\documentclass[a4paper,fleqn,usenatbib]{mnras_arxiv}

\usepackage[T1]{fontenc}
\usepackage{ae,aecompl}

\usepackage{graphicx}	
\usepackage{amsmath}	
\usepackage{amssymb}	
\usepackage{color,soul} 

\title[polarization of HD~189733]{The polarization of HD~189733}

\author[K. Bott et al.]{Kimberly Bott$^{1,2}$\thanks{E-mail: k.bott@unsw.edu.au (KB)},
Jeremy Bailey,$^{1,2}$,
Lucyna Kedziora-Chudczer$^{1,2}$,
\newauthor
Daniel V. Cotton$^{1, 2}$,
P.W. Lucas$^{3}$,
Jonathan P. Marshall$^{1, 2}$
J.H. Hough$^{3}$
\\
$^{1}$School of Physics, UNSW Australia, NSW 2052, Australia\\
$^{2}$Australian Centre for Astrobiology, UNSW Australia, NSW 2052, Australia.\\
$^{3}$Centre for Astrophysics Research, Science and Technology Research Institute, University of
Hertfordshire, Hatfield AL10 9AB, UK}

\date{Accepted XXX. Received YYY; in original form ZZZ}

\pubyear{2016}

\begin{document}
\label{firstpage}
\pagerange{\pageref{firstpage}--\pageref{lastpage}}
\maketitle

\begin{abstract}
We present linear polarization observations of the exoplanet system HD 189733 made with the HIgh
Precision Polarimetric Instrument (HIPPI) on the Anglo-Australian Telescope (AAT). The observations have
higher precision than any previously reported for this object. They do not show the large amplitude polarization
variations reported by \citet{berdyugina08,berdyugina11}. Our results are consistent with polarization data presented by \citet{wiktorowicz15b}.
A formal least squares fit of a Rayleigh-Lambert model yields a polarization amplitude of 29.4
$\pm$ 15.6 parts-per-million. We observe a background constant level of polarization of $\sim$
55--70 ppm, which is a little higher than expected for interstellar polarization at the distance of HD 189733. 
\end{abstract}

\begin{keywords}
polarization -- techniques: polarimetric -- planets and satellites: atmospheres -- planets and satellites: individual: HD 189733b
\end{keywords}



\section{Introduction}

The characterisation of exoplanet atmospheres, particularly with ground based telescopes, is a difficult task. Polarimetry offers a useful approach, providing a strong contrast between the star and planet as the star's light is typically unpolarized \citep{seager00}.
Large and tightly orbiting hot Jupiter planets scatter enough light to  potentially produce a polarisation signal dependent largely upon the composition of the atmosphere \citep{hough03}.
The detection of linearly polarized light from an exoplanet system can provide information about a planet's orbital
orientation and about the properties of the particles that scatter the light in its atmosphere. The
technique therefore provides complementary information to other characterisation techniques such as
transit and eclipse spectroscopy.

\citet{seager00} modelled the expected polarization levels
for hot Jupiter-type systems and predicted that linear polarization
varying over the orbital cycle at the tens of parts-per-million level
might be present in the combined light of the star and planet. Since
Sun-like stars are thought to generally have low polarizations \citep{bailey10, cotton16} this technique 
provides good contrast
and should be achievable from ground-based observations.  Past attempts at polarized
light detection such as those described in \citet{lucas09} have not detected significant
planetary polarization signals. 

The HD 189733 hot Jupiter system \citep{bouchy05, torres08} is one of the brightest and best studied transiting exoplanet systems.
Observations of transits and eclipses of HD 189733b using the Hubble Space Telescope have shown strong 
evidence for a Rayleigh scattering haze in its atmosphere, which shows up as increasing absorption
towards shorter wavelengths in the transit spectrum \citep{pont08,pont13} as well as reflected light
seen in the secondary eclipse at blue wavelengths \citep{evans13}. This makes HD 189733b a promising
target for polarimetric observations.

\vspace{-3pt}
However, polarization observations of this system to date have produced conflicting results.
\citet{berdyugina08} reported an unexpectedly high level of polarization variation of $\sim$200 ppm
(parts-per-million) from the HD 189733
exoplanet system using the DIPol instrument \citep{piirola05} at La Palma's KVA (Royal Swedish Academy of Sciences) telescope. Subsequently
\citet{wiktorowicz09} reported no polarization variation with an upper limit of 79 ppm 
(99 per cent confidence) 
using the POLISH instrument. Further observations \citep{berdyugina11} using TurPol \citep{piirola73} on
the Nordic Optical Telescope in three bands showed a polarization signal in the \textit{U} and \textit{B}
bands but not in the \textit{V} band. They argued that the non-detection by \citet{wiktorowicz09} was due
to the use of too red a band. The amplitude reported in the \textit{B} band was 100 $\pm$ 10 ppm. The polarization was attributed 
to Rayleigh scattering from the planet's atmosphere. 

More recently \citet{wiktorowicz15b} have reported \textit{B} band polarization measurements of HD 189733
using the POLISH2 instrument on the Lick Observatory 3-m telescope. They set a limit on the \textit{B} band
polarization amplitude of 60 ppm (99.7 per cent confidence). 

This paper presents the most sensitive measurements of the system to date using a blue broadband 
filter in an attempt to distinguish between these conflicting results.


\vspace{-10pt}

\section{Observations}

HD 189733 was observed during three observing runs on the 3.9-m Anglo-Australian Telescope (AAT) at Siding
Spring Observatory, New South Wales, Australia. The observations were made with the HIgh Precision
Polarimetric Instrument (HIPPI)\citep{bailey15}. The dates of observations were 2014 Aug 28-31, 2015 May 23,
26 and Jun 26. HIPPI is an aperture polarimeter using a ferroelectric liquid crystal (FLC) modulator, a
Wollaston prism analyser and two photomultiplier tubes (PMT) as detectors. The FLC provides a 500 Hz
primary modulation which is used together with two additional stages of slower modulation obtained by
rotation of the Wollaston prism and detectors, and finally by rotating the whole instrument to four
position angles (0, 45, 90, 135 degress) using the AAT's Cassegrain rotator. 

HIPPI has been shown from repeat observations of bright low polarization stars to deliver a precision
of 4.3 $\times$ 10$^{-6}$ (4.3 parts-per-million or ppm) or better \citep{bailey15}. This is
comparable to or better than the precisions reported from polarimeters based on photoelastic
modulators such as the Pine Mountain Observatory polarimeter \citep{kemp81} , PlanetPOL \citep{hough06}, POLISH \citep{wiktorowicz08} and POLISH2
\citep{wiktorowicz15a,wiktorowicz15b}.

HIPPI uses Hamamatsu H10720-210 PMT modules that have ultra-bialkali photocathodes with a peak
quantum efficiency of 43 per cent at 400 nm. The HD 189733 observations were made through a 500 nm short
pass filter (referred to as 500SP) to exclude any red light from the K 1.5 V star. The overall bandpass covers 350 to 500 nm (the polarization optics cut off light below 350 nm) and the effective wavelength
calculated from the bandpass model described by \citet{bailey15} is 446.1 nm. This is a similar but
somewhat broader band than the \textit{B} band used by \cite{berdyugina11} and \cite{wiktorowicz15b}.

The moonlit sky can contribute a significant polarized background signal within our 6.7 \arcsec diameter aperture. The 2014 August observations were all made with lunar illumination phases less than
35 per cent but with the Moon near 90$^\circ$ seperation from HD 189733b where Rayleigh scattering is at a maximum. The 2015 observations were all made after the Moon had set. Any residual background signal
is subtracted from the data using a sky observation made immediately after each science observation at each Cassegrain rotator position.

The telescope introduces a small polarization (telescope polarization or TP) that must be corrected for. As described
in \citet{bailey15} we determine the TP using observations of a number of stars we believe to have very low polarization
either based on previous PlanetPol observations \citep{hough06,bailey10} or because of their small distances and
expected low levels of interstellar polarization \citep[see ][]{bailey10,cotton16}. The telescope polarization has
been found to be stable during each run, and between the 2015 May and June runs, but changed signficantly from 2014
to 2015, probably as a result of the realuminization of the AAT primary mirror in early 2015. Determinations of TP in
the SDSS g$^\prime$ filter are reported in \citet{bailey15} and \cite{cotton16} giving values of 48 $\pm$ 5 ppm in 2014, 
and 36 $\pm$ 1 ppm in 2015.

We have also made a smaller number of TP observations directly in the 500SP filter used with HD~189733
as listed in Table \ref{tab_tp}. These observations consistently show telescope polarizations of
22 to 25 per cent higher values than the more extensive $g^\prime$ band observations.

\begin{table}
\caption{Low polarization star measurements to determine telescope polarization (TP) in the 500SP filter. The adopted TP for the August 2014 and May 2015 runs use only measurements acquired on that run. The June 2015 run uses measurements from both May and June 2015.}
\begin{flushleft}
\begin{tabular}{llcll}
\hline
Star &  Date &     $p$ (ppm) &   \hspace{2 mm}$\theta$ ($\degr$) \\
\hline
$\beta$ Hyi  &    29 Aug &   60.7 $\pm$ \hspace{1.5 mm}4.4 & 109.3 $\pm$ \hspace{1.5 mm}2.1    \\
$\beta$ Hyi        &    30 Aug &   63.6 $\pm$ \hspace{1.5 mm}4.3 & 110.0 $\pm$ \hspace{1.5 mm}2.0    \\
BS 5854 &    31 Aug &   60.6 $\pm$ \hspace{1.5 mm}4.7 & 111.4 $\pm$ \hspace{1.5 mm}2.2    \\
$\beta$ Hyi        &    31 Aug &   51.9 $\pm$ \hspace{1.5 mm}4.6 & 112.4 $\pm$ \hspace{1.5 mm}2.6    \\
\hline
Adopted TP & Aug 2014 & 59.1 $\pm$ \hspace{1.5 mm}2.2 & 110.7 $\pm$ \hspace{1.5 mm}1.1    \\
\hline
BS 5854  &    22 May &   38.7 $\pm$ \hspace{1.5 mm}5.3 & \hspace{1.5 mm}83.3 $\pm$ \hspace{1.5 mm}3.9    \\
Sirius   &    23 May &   44.8 $\pm$ \hspace{1.5 mm}0.9 & \hspace{1.5 mm}87.2 $\pm$ \hspace{1.5 mm}0.6    \\
$\beta$ Hyi &    26 May &   49.8 $\pm$ 11.7 & \hspace{1.5 mm}88.4 $\pm$ \hspace{1.5 mm}6.9    \\
\hline
Adopted TP & May 2015 & 44.3 $\pm$ \hspace{1.5 mm}3.4 & \hspace{1.5 mm}86.5 $\pm$ \hspace{1.5 mm}2.3    \\
\hline
BS 5854 &   26 Jun  &   50.2 $\pm$ \hspace{1.5 mm}4.8 & \hspace{1.5 mm}97.2 $\pm$ \hspace{1.5 mm}2.8    \\
\hline 
Adopted TP & Jun 2015 & 45.8 $\pm$ \hspace{1.5 mm}2.8 & \hspace{1.5 mm}86.7 $\pm$ \hspace{1.5 mm}1.8    \\
\hline
\hline
\end{tabular}
\end{flushleft}
\label{tab_tp}
\end{table} 

Full details of the observation, calibration, and data reduction procedures with HIPPI can be found in \citet{bailey15}.









\vspace{-10pt}

\section{Results}

The unbinned polarization results before efficiency correction on HD 189733, corrected for telescope polarization using the values in Table
\ref{tab_tp}, are listed in
Table \ref{tab_results}. Each of these measurements is the result of approximately one hour of
total observation (half this for the last two points on 2014:08:31). We list the mid-point time and
the normalised
Stokes parameters Q/I and U/I are given in parts-per-million (ppm) on the equatorial system. 
One further correction is required to the data and this is for the wavelegth dependent
modulation efficiency of the instrument. This is calculated using the bandpass model described by
\citet{bailey15} specifically for each observation. The value is close to 82.4 per cent for
all of these observations. 

The nightly means of the corrected observations each consisting of approximately two to three hours
total observing time are given in Table \ref{tab_binned_results}. One point (2014:08:31 12:47:49)
has been omitted from the binned data as it is believed to be affected by variable sky background due to the Moon setting mid observation.

\begin{table}
\caption{The data before binning and efficiency correction. The datum marked with a $\dagger$ is believed to be a physical outlier and is not binned with the other data.  }
\begin{tabular}{lll}
\hline UT Date and Time & Q/I (ppm) &   U/I (ppm)   \\   \hline
2014:08:28 11:25:52                  & 64.2 $\pm$ 14.0  & 11.5 $\pm$ 20.2 \\
2014:08:28 12:29:38                  & -14.3 $\pm$ 22.3 & 18.0 $\pm$ 21.2 \\
2014:08:28 13:31:19                  & 28.3 $\pm$ 21.6  & 54.0 $\pm$ 22.3 \\
2014:08:29 10:18:49                  & 11.9 $\pm$ 19.2  & 29.0 $\pm$ 19.0 \\
2014:08:29 11:25:24                  &  \phantom{0}9.9  $\pm$ 19.0  & -8.1 $\pm$ 19.1 \\
2014:08:29 12:29:08                  & 30.0 $\pm$ 19.3  & 29.9 $\pm$ 20.7 \\
2014:08:30 10:53:57                  & 39.5 $\pm$ 21.1  & 14.9 $\pm$ 21.8 \\
2014:08:30 12:00:11                  & 38.9 $\pm$ 19.2  & 15.1 $\pm$ 21.8 \\
2014:08:30 13:05:23                  & 51.5 $\pm$ 19.5  & 21.4 $\pm$ 19.3 \\
2014:08:31 10:18:26                  & 58.0 $\pm$ 18.6  & 17.4 $\pm$ 19.0 \\
2014:08:31 11:24:40                  & 33.1 $\pm$ 19.0  & 44.4 $\pm$ 18.4 \\
2014:08:31 12:11:13                  & 42.4 $\pm$ 25.0  & 72.3 $\pm$ 25.0 \\
2014:08:31 12:47:49 $\dagger$        & 77.0 $\pm$ 26.7  & 85.2 $\pm$ 26.6 \\
2015:05:23 16:50:38                  & 22.1 $\pm$ 22.8  &  \phantom{0}5.1  $\pm$ 23.1 \\
2015:05:23 17:50:59                  & 45.0 $\pm$ 20.9  & 40.1 $\pm$ 20.9 \\
2015:05:23 18:50:20                  & 38.7 $\pm$ 20.6  & 18.6 $\pm$ 20.9 \\
2015:05:26 16:30:20                  & 54.0 $\pm$ 19.7  & 21.6 $\pm$ 19.7 \\
2015:05:26 17:32:55                  & 57.4 $\pm$ 19.1  & 47.9 $\pm$ 19.3 \\
2015:05:26 18:32:07                  & 69.5 $\pm$ 21.2  & 28.5 $\pm$ 21.4 \\
2015:06:26 16:33:18                  & -6.2 $\pm$ 18.8  & 45.5 $\pm$ 18.9 \\
2015:06:26 17:30:07                  & 37.4 $\pm$ 19.3  & 63.7 $\pm$ 19.2 \\

\hline
\end{tabular}
\label{tab_results}
\end{table}

\begin{table*}
\caption{Nightly means of the linear polarization of HD 189733.} 
\begin{tabular}{lllll}
\hline UT Date and Time & HMJD & Phase & Q/I (ppm) &   U/I (ppm)  \\  \hline
2014:08:28 12:28:56               & 56897.52397  & 0.30255 & 31.7 $\pm$ 13.8 & 33.8 $\pm$ 14.9  \\
2014:08:29 11:24:27               & 56898.47916  & 0.73309 & 21.0 $\pm$ 13.4 & 20.6 $\pm$ 13.8 \\
2014:08:30 11:59:50               & 56899.50369  & 0.19488 & 52.6 $\pm$ 14.0 & 20.8 $\pm$ 14.3  \\
2014:08:31 11:07:29               & 56900.46731  & 0.62923 & 54.5 $\pm$ 14.3 & 47.6 $\pm$ 14.2  \\
2015:05:23 17:50:39               & 57165.74522  & 0.20064 & 42.8 $\pm$ 15.1 & 25.8 $\pm$ 15.2  \\
2015:05:26 17:31:47               & 57168.73232  & 0.54704 & 73.2 $\pm$ 14.0 & 39.7 $\pm$ 14.1  \\
2015:06:26 17:01:43               & 57199.71313  & 0.51133  & 19.0 $\pm$ 16.4 & 66.3 $\pm$ 16.4  \\
\hline
Average                                &                        &                & 42.7 $\pm$ 5.4    & 35.4 $\pm$ 5.5 \\
\hline
\end{tabular}
\label{tab_binned_results}
\end{table*}

Orbital phase is calculated according to the ephemeris \citep{triaud09} where zero phase corresponds
to mid-transit.

\begin{equation}
\mathrm{ T} = \mathrm{ HMJD} \, 53988.30339 + 2.21857312 \, \mathrm{ E}  .
\end{equation} 

The errors of our nightly means are typically 13--16 ppm. This can be compared with errors of typically
20--40 ppm for the nightly means of POLISH2 observations \citep{wiktorowicz15b}


\vspace{-10pt}

\section{Discussion}

\vspace{-3pt}

\subsection{Comparison with previous results}
\label{sec_results}

Our results differ significantly from those reported by \citet{berdyugina11}. While we see consistently
positive values of Q/I and U/I, \citet{berdyugina11} show near zero values at phases 0.0 and 0.5, with a
strong negative excursion around phases 0.3 and 0.7, reaching an amplitude of nearly 100 ppm in Q/I. Even
if we allow an arbitrary zero point shift, our data are not consistent with such a large amplitude
variation in Q/I.

Our data are in much better agreement with the results of \citet{wiktorowicz15b}. This dataset shows
generally positive Q/I and U/I with average values (from data in their Table 3) of Q/I = 19.2 $\pm$ 4.1 ppm and U/I = 40.3 $\pm$ 3.5
ppm, which are in reasonable agreement with our averages of Q/I = 42.7 $\pm$ 5.4 and U/I = 35.4 $\pm$ 5.5
ppm. The differences can probably be understood as due to uncertainties in the telescope polarization of
the Lick telescope which \citet{wiktorowicz15b} report is variable at the 10 ppm level.  

\begin{figure}
	\includegraphics[angle=0,width=\columnwidth]{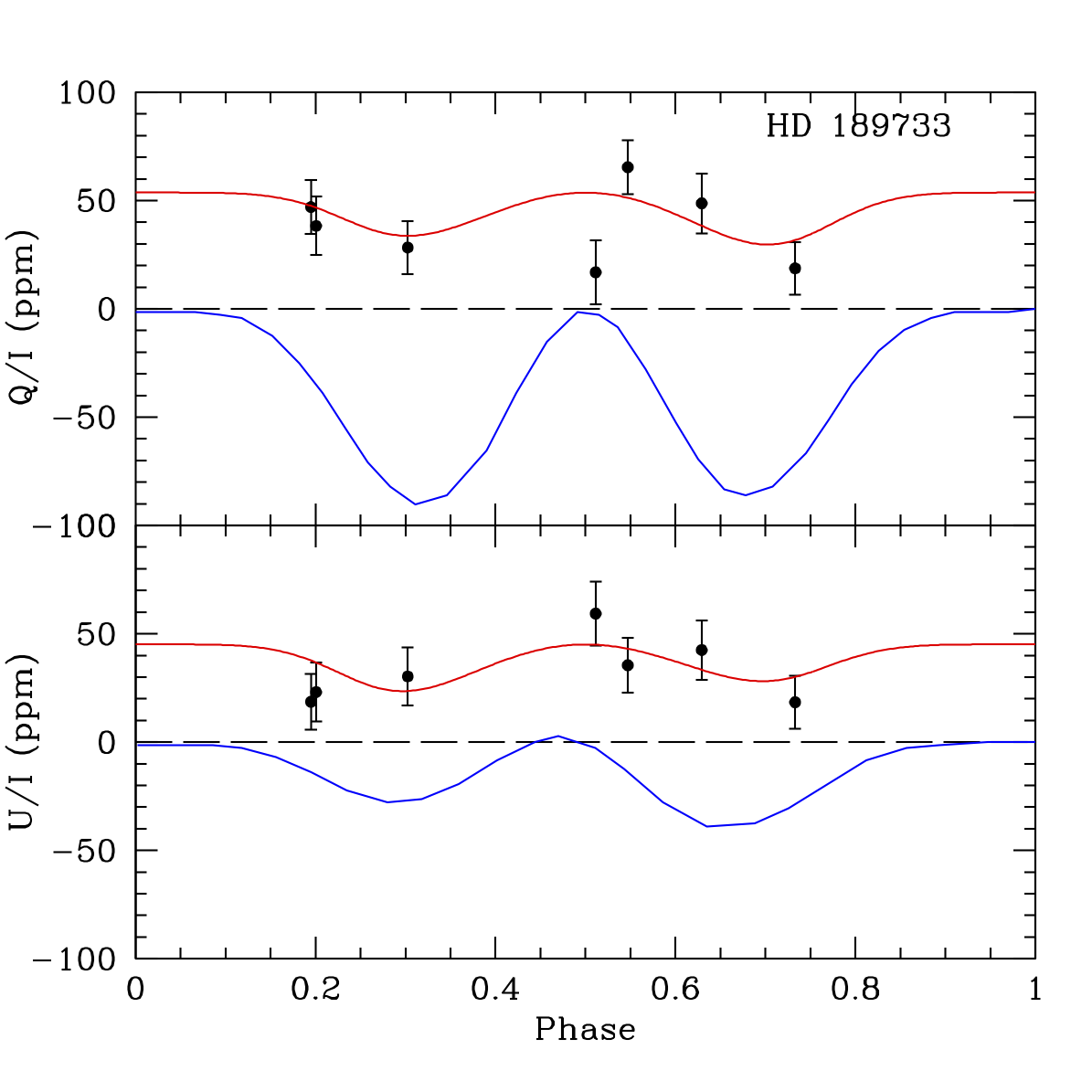}
    \caption{HIPPI measurements of the polarized light from the HD 189733 system. The binned
    measurements from Table \ref{tab_binned_results} are shown. The red curve is a least sqares fit
    of a Rayleigh-Lambert model as described in Section \ref{sec_ray_lamb}. The blue lines   
    show the best-fit curves for the data of \citet{berdyugina11}} 
    \label{fig_hd189}
\end{figure}


\vspace{-5pt}

\subsection{Rayleigh-Lambert model}
\label{sec_ray_lamb}

We have fitted our data with a Rayleigh-Lambert model for the expected polarization variation. This
is a simple analytic model which calculates the intensity according to the expected phase variations 
for a Lambert sphere, and assumes the polarization follows the phase function for Rayleigh scattering (see: \citet{seager00} and \citet{wiktorowicz09}). To find the best fit we use a
Levenberg-Marquardt non-linear least squares algorithm \citep{press92} with five parameters: the polarization zero point offsets in Q/I and U/I ($Z_q$ and $Z_u$), the polarization
amplitude $p$ which allows the effects of depolarisation processes such as multiple scattering to be taken into account, the position angle of the major axis of the projected orbit ellipse on the sky $P\!A$ and
the orbital inclination $i$. The fitted parameters and their uncertainties (determined from the covariance matrix
of the fit) are listed in Table \ref{tab_fit}. The fitted model is shown by the red curve on Figure \ref{fig_hd189}.

\begin{table}
\centering
\caption{Parameters of Rayleigh-Lambert fit to HIPPI linear polarization observations of HD 189733}
\begin{tabular}{lll}
\hline
Parameter & Value & units \\
\hline
$Z_q$ & 53.7 $\pm$ 9.6& ppm \\
$Z_u$ & 45.1 $\pm$ 9.5 & ppm \\
$p$ & 29.4 $\pm$ 15.6 & ppm \\
$pa$ & 114.4 $\pm$ 19.0 & degrees \\
$i$ & 87.6 $\pm$ 6.6 & degrees \\
\hline
\end{tabular}
\label{tab_fit}
\end{table}

The less than 2-sigma uncertainty on the polarization amplitude means that this cannot be considered as a
detection of polarized light from the planet. A model with no polarization variation due to the planet is
still an accepteble fit to the data. However, it is interesting that our best fit polarization amplitude
of 29.4 $\pm$ 15.6 ppm is at a level that is in agreement with plausible values for Rayleigh scattering 
from the planet. For example \citet{lucas09}  estimate an amplitude of 26 ppm for a Rayleigh-like multiple
scattering model with a single scattering albedo of 0.99. This is
an optimistic model because in reality we would expect a modest reduction in the amplitude due to (i) atomic and molecular absorption
features and (ii) the Rayleigh depolarisation factor of 0.02 \citep{penndorf57, hansen74} which reduces the polarisation at each scattering event.  Considering the geometric albedo from a fractional eclipse depth of $\sim 126$ ppm from 290-450 nm reported in \citet{evans13}, and the Rayleigh scattering model grid of \citet{buenzli09} which would place an upper limit to the amplitude of the polarised light contribution at about 30 per cent of the geometric albedo, we can estimate an upper bound for a Rayleigh scattering atmosphere to be $\sim 37.8 ppm$.

A more extensive set of observations with an instrument like HIPPI on a 4-m class
telescope might be capable of detecting the expected planetary signal in this and other bright hot
Jupiter systems. Putting a similar instrument on an 8-m class telescope should enable the measurement
uncertainties to be halved from $\sim$ 14 ppm to $\sim$ 7 ppm, making detection of the expected signals
at levels of $\sim$ 20 ppm possible.

\vspace{-10pt}

\subsection{Constant polarization}

A clear result of our observations, is that in addition to any polarization from the planet, there is a
constant background level of polarization from HD 189733. Depending on what we assume about the planetary
signal, this ranges from around 55 ppm, if we use the values given in Section \ref{sec_results} for no
planetary signal, to 70 ppm for the best fit model of Table \ref{tab_fit}. Interstellar polarization is 
the most likely source of polarization in typical solar-like stars
\citep{bailey10, cotton16}. 

Extrapolating from the trends in Figure 4 in \citet{cotton16}, we would expect a star at $\sim 19$ pc \citep{koen10}, in the Northern Hemisphere and in the galactic plane to have a polarization between 15--40 ppm. 
This is rather less than the values we find for HD 189733.

Circumstellar dust can also produce a constant offset for the polarized light signal from a system. However, \citet{bryden09} found HD~189733 was unlikely to have circumstellar dust substantial enough to affect polarisation measurements based upon measurements of infrared excess. The effect of circumstellar dust on a system's polarized light measurements would be an addition to the offsets $Z_q$ and $Z_u$ which would not vary over the timescale of the planet's orbit.

\vspace{-10pt}

\subsection{Effect of starspots on system polarimetry}
There are other possible sources that could be contributing to polarized light in the system.
HD 189733A is an active BY Draconis type variable star: a star whose brightness varies due to 
star spots moving across its surface. Star spots can cause linear polarization by breaking the circular
symmetry of the limb polarization \citep{moutou07}. However \citet{berdyugina11} found that starspots would only cause a maximum of 3 ppm contribution to polarized light based on the photometric transit curve deviations from \citet{winn07}. Similarly, \citet{kostogryz15} estimated the contribution of starspots on the polarized light signal to be only  $\sim 2 \times 10^{-6}$. The interplay between the planet and starspot symmetry breaking can produce detectable effects during the planet's transit. However none of our observations are taken during transit. 
 
 The linearly polarized light from the starspots themselves, is negligible \citep{afram15} under most circumstances. Magneto-optical effects such as the transverse Zeeman effect \citep{huovelin91} or the Faraday effect \citep{calamai75} could introduce noise to the linear polarisation measuements if enough lines were present within our bandpass or if significant starspot coverage is present along with extended ionised gases from the planet's atmosphere respectively.

The rotational period of HD 189733A is known to be longer, at 11.8 days \citep{moutou07}, than the orbital period of HD 189733b. With additional observations it should be possible to determine whether the magnitude of activity associated with starspots is significant, and disentangle it from that due to reflection from clouds in the planetary atmosphere. Regardless, in any analysis focused on the orbital period, starspot effects will average out over time.


\vspace{-10pt}

\subsection{Effect of Saharan dust on La Palma observations}

The large polarization amplitudes for HD 189733 reported by \citet{berdyugina11} are not seen in
the other three datasets now reported for this object, \citet{wiktorowicz09}, \citet{wiktorowicz15b} 
and this work.

The strongest evidence for polarization variation found by \citet{berdyugina11} was in the \textit{B} band 
TurPol data from the Nordic Optical Telescope on La Palma obtained on dates of 2008 Apr 18-24, and 2008
Aug 2-9. They note that the August data was affected by Saharan dust but argue that this should not cause
any effect on the results as ``\textit{TurPol enables exact compensation of any background polarization that is
not variable ...}''. This statement misunderstands how airborne dust impacts on precision polarimetry. It
is not a background sky polarization that could be automatically subtracted by the instrument's
sky-subtraction capability. The dust introduces a spurious polarization into starlight passing through
the dust. The effects were studied in detail by \citet{bailey08} using observations of a Saharan dust
event observed with PlanetPol in 2005.

Data from the Carlsberg Meridian Telescope
\footnote{http://www.ast.cam.ac.uk/\raisebox{-0.5ex}{\textasciitilde}dwe/SRF/camc\_extinction.html} on La Palma show a substantial Saharan dust event
over 2008 Aug 2 to 9 with the \textit{$r^\prime$} band extinction ranging from 0.212 to 0.377 throughout this period
(compared with a normal clear sky level of $\sim$0.09). This
is a larger and more extended event than the one observed in 2005 by \citet{bailey08} which led to
spurious polarization effects up to 48 ppm at 56\degr \; zenith distance. It seems highly likely that the
HD~189733 data reported by \citet{berdyugina11} are significantly affected by this dust. As the individual
observations are not reported, it is not possible to judge the extent of the problem. However, without 
exclusion or correction of affected data, this cannot be regarded as a reliable dataset for precision
polarimetry.

\vspace{-15pt}

\section{Conclusions}

We have reported new polarization observations of the hot Jupiter exoplanet system HD 189733. The
observations have higher precision than any previously reported for this object. We do not detect the
large polarization amplitude (100-200 ppm) planetary signals previously reported
\citep{berdyugina08,berdyugina11}. Our results agree reasonably well with the results of
\citet{wiktorowicz15b} showing generally positive polarization values in Q/I and U/I, and at most
a small planetary polarization signal.

A least-squares fit of a Rayleigh-Lambert model gives a polarization amplitude of 29.4 $\pm$ 15.6 ppm.
While this signal has less than 2-sigma significance and cannot be claimed as a detection of planetary
polarization, it is at a level consistent with a plausible polarization amplitude from the
planet. It is consistent with a multiply-scattering atmosphere which could produce the albedo measurements taken by \citet{evans13}. This suggests that a more extensive series of observations, or observations on a larger telescope
should enable the planetary polarization to be detected and measured.

HD 189733 has a significant constant background level of polarization that is a somewhat higher than would
be expected for interstellar polarization. This could be due to the non-unformity of the Local Hot Bubble interstellar medium.

We suggest that the polarization data of \citet{berdyugina11}, which shows a large polarization amplitude
inconsistent with that reported by other groups, may be unreliable as a result of spurious polarization
due to a Saharan dust event over the La Palma observatory in August 2008.

\vspace{-10pt}

\section*{Acknowledgements}

The development of HIPPI was funded by the Australian Research Council through Discovery Projects
grant DP140100121 and by the UNSW Faculty of Science through its Faculty Research Grants program. JPM is supported by a UNSW Vice-Chancellor's Fellowship. The
authors thank the Director and staff of the Australian Astronomical Observatory for their advice and 
support with interfacing HIPPI to the AAT and during the three observing runs on the telescope. The authors wish to thank referee Hans Martin Schmid for constructive criticism of the paper.




\vspace{-10pt}








\bsp	
\label{lastpage}
\end{document}